\documentclass[conference]{IEEEtran}
\IEEEoverridecommandlockouts

\usepackage{cite}
\usepackage{amsmath,amssymb,amsfonts}
\usepackage{amsthm}
\usepackage{stmaryrd}
\usepackage{algorithmic}
\usepackage{graphicx}
\usepackage{textcomp}
\usepackage{xcolor}
\usepackage{multirow}
\usepackage{enumitem}
\usepackage{comment}
\usepackage{tikz}
\usepackage{pgfplots}
\usetikzlibrary{pgfplots.dateplot}
\usepackage{pgfplotstable}
\usepackage{filecontents}

\newtheorem{example}{Example}

\newcommand{\verisol}{\textsc{VeriSol}}
\newcommand{\corral}{\textsc{Corral}}

\usepackage{booktabs}
\usepackage{color}
\usepackage{xcolor}
\usepackage{float}
\usepackage{listings}
\usepackage{hyperref}
\usepackage{amssymb}
\usepackage{amsmath}
\usepackage{xspace}
\usepackage{graphicx}
\usepackage{multirow}
\usepackage{url}
\usepackage[T1]{fontenc}
\usepackage{beramono}
\usepackage{tikz}
\usepackage{balance}
\usepackage[utf8]{inputenc}
\usepackage[T1]{fontenc}
\usepackage{microtype}

\newcommand{\Space}[1]{}

\newcommand{\Shuvendu}[1]{\textcolor{blue}{[SKL:#1]}}

\newcommand{\Def}[2]{\expandafter\newcommand\csname rmk-#1\endcsname{#2}}
\newcommand{\Use}[1]{\csname rmk-#1\endcsname}

\newcommand{\set}[1]{\ensuremath{\{#1\}}}

\definecolor{light}{gray}{0.80}

\newcommand{\proc}[1]{{\textit{f}_{#1}}}
\newcommand{\pproc}{\proc{}} 

\newcommand{\Stmts}{\textit{Stmts}}

\newcommand{\stmt}[1]{\textit{st}_{#1}}
\newcommand{\sstmt}{\stmt{}} 

\newcommand{\solStmts}{\textit{SolStmts}}
\newcommand{\solStmt}[1]{\textit{sst}_{#1}}
\newcommand{\solRequire}{\texttt{require}}
\newcommand{\solAssert}{\texttt{assert}}
\newcommand{\solIfelse}[3]{\texttt{if} \ (#1) \ \{  #2   \} \ \texttt{else} \ \{   #3 \} }
\newcommand{\solWhilen}{\texttt{while}}
\newcommand{\solWhile}[2]{\solWhilen \ (#1) \ \texttt{do} \ \{ #2 \}}

\newcommand{\Exprs}{\textit{Exprs}}
\newcommand{\expr}[1]{\textit{e}_{#1}}
\newcommand{\bexpr}{\expr{}}

\newcommand{\solExprs}{\textit{SolExprs}}
\newcommand{\solExpr}[1]{\textit{se}_{#1}}

\newcommand{\bfont}[1]{\mathbf{#1}}

\newcommand{\bvar}[1]{\texttt{#1}}
\newcommand{\varn}[1]{\bvar{#1}}
\newcommand{\varnVector}[1]{\vec{\bvar{#1}}}

\newcommand{\bint}{\bfont{int}}
\newcommand{\bRef}{\bfont{Ref}}

\newcommand{\bop}[1]{\bfont{op}(#1)}

\newcommand{\bskip}{\bfont{skip}}
\newcommand{\bcalln}{\bfont{call}}
\newcommand{\bcall}[3]{{\bcalln} \ {#3} := #1(#2)}

\newcommand{\bifelse}[3]{\bfont{if} \ (#1) \ \{  #2   \} \ \bfont{else} \ \{   #3 \} }

\newcommand{\bwhile}[2]{\bfont{while} \ (#1) \ \bfont{do} \ \{ #2 \}}

\newcommand{\bhavoc}[1]{\bfont{havoc} \ {#1}}

\newcommand{\bassertn}{\bfont{assert}}
\newcommand{\bassumen}{\bfont{assume}}

\newcommand{\bassign}[2]{{#1} := {#2}}

\newcommand{\bassert}[1]{\bassertn \ #1}
\newcommand{\bassume}[1]{\bassumen \ #1}

\newcommand{\Seq}[2]{{#1};{#2}}

\newcommand{\arrSel}[2]{{#1}[{#2}]}

\lstdefinelanguage{boogie}{
  keywords={%
    proc,free, function,returns,var,int,bool,call,return,assert,assume,goto,havoc,modifies,requires,ensures,while,if,const,axiom,foreach
  },
  morecomment=[l]{//},
  basicstyle=\ttfamily\footnotesize,
  escapechar=\%
}

\definecolor{impacted}{rgb}{0.7,0.7,0.7}
\definecolor{syntchanged}{rgb}{1.,0.75,0.75}
\definecolor{linenumbergray}{rgb}{0.5,0.5,0.5}
\definecolor{gitdel}{RGB}{255,236,236}
\definecolor{gitdelfocused}{RGB}{248,203,203}
\definecolor{gitadd}{RGB}{234,255,234}
\definecolor{gitaddfocused}{RGB}{166,235,166}

\makeatletter
\newenvironment{btHighlight}[1][]
{\begingroup\tikzset{bt@Highlight@par/.style={#1}}\begin{lrbox}{\@tempboxa}}
{\end{lrbox}\bt@HL@box[bt@Highlight@par]{\@tempboxa}{impacted}\endgroup}

\newcommand\btHL[1][]{%
  \begin{btHighlight}[#1]\bgroup\aftergroup\bt@HL@endenv%
}
\def\bt@HL@endenv{%
  \end{btHighlight}%
  \egroup
}
\newcommand{\bt@HL@box}[3][]{%
  \tikz[#1]{%
    \pgfpathrectangle{\pgfpoint{1pt}{0pt}}{\pgfpoint{\wd #2}{\ht #2}}%
    \pgfusepath{use as bounding box}%
    \node[anchor=base west, fill=#3, outer sep=0pt,inner xsep=0pt, inner ysep=-0.5pt, #1]{\raisebox{1pt}{\strut}\strut\usebox{#2}};
  }%
}
\makeatother

\lstdefinestyle{C-github}{basicstyle=\ttfamily\small,
        xleftmargin=5.0ex,
        language=C,
        numbers=left,
        numberstyle=\tiny\color{linenumbergray},
        moredelim=**[is][\btHL]{@i@}{@i@},
        moredelim=**[is][{\btHL[fill=gitdel]}]{@d@}{@d@},
        moredelim=**[is][{\btHL[fill=gitdelfocused, rounded corners=2pt]}]{@df@}{@df@},
        moredelim=**[is][{\btHL[fill=gitadd]}]{@a@}{@a@},
        moredelim=**[is][{\btHL[fill=gitaddfocused, rounded corners=2pt]}]{@af@}{@af@},
        escapeinside={(*@}{@*)}}

\newcommand{\Contract}{\mathcal{C}}
\newcommand{\Policy}{\pi}

\begin{document}
\title{Formal Specification and Verification of Smart Contracts in Azure Blockchain}


\author{\IEEEauthorblockN{Yuepeng Wang}
\IEEEauthorblockA{\textit{University of Texas at Austin} \\
ypwang@cs.utexas.edu}
\and
\IEEEauthorblockN{Shuvendu K. Lahiri}
\IEEEauthorblockA{\textit{Microsoft Research} \\
shuvendu@microsoft.com}
\and
\IEEEauthorblockN{Shuo Chen}
\IEEEauthorblockA{\textit{Microsoft Research} \\
shuochen@microsoft.com}
\and
\IEEEauthorblockN{Rong Pan}
\IEEEauthorblockA{\textit{University of Texas at Austin} \\
rpan@cs.utexas.edu}
\and
\IEEEauthorblockN{\ \ \ \ \ \ \ \ \ \ Isil Dillig}
\IEEEauthorblockA{\ \ \ \ \ \ \ \ \ \ \textit{University of Texas at Austin} \\
\ \ \ \ \ \ \ \ \ \ isil@cs.utexas.edu}
\and
\IEEEauthorblockN{Cody Born}
\IEEEauthorblockA{\textit{Microsoft Azure} \\
coborn@microsoft.com}
\and
\IEEEauthorblockN{Immad Naseer}
\IEEEauthorblockA{\textit{Microsoft Azure} \\
imnaseer@microsoft.com}
}

\maketitle

\begin{abstract}
Ensuring correctness of smart contracts is paramount to ensuring trust in  blockchain-based systems. 
This paper studies the safety and security of smart contracts in the \emph{Azure Blockchain Workbench}, an enterprise Blockchain-as-a-Service offering from Microsoft. 
As part of this study, we formalize \emph{semantic conformance} of smart contracts against a state machine model with access-control policy and develop a highly-automated formal verifier for Solidity that can produce proofs as well as counterexamples. 
We have applied our verifier {\sc VeriSol} to analyze {\it all} contracts shipped with the Azure Blockchain Workbench, which includes application samples as well as a governance contract for Proof of Authority (PoA). 
We have found previously unknown bugs in these published smart contracts.
After fixing these bugs, {\sc VeriSol} was able to successfully perform full verification for all of these contracts. 
\end{abstract}

\section{Introduction}\label{sec:intro}

As a decentralized and distributed consensus protocol to maintain and secure a shared ledger, the  blockchain  is seen as a disruptive technology with far-reaching impact on diverse areas. 
As a result, major cloud platform companies, including Microsoft, IBM, Amazon, SAP, and Oracle, are offering Blockchain-as-a-Service (BaaS) solutions, primarily targeting enterprise scenarios, such as financial services, supply chains, escrow, and consortium governance. 
A recent study by Gartner predicts that the business value-add of the blockchain has the potential to exceed \$3.1 trillion by 2030~\cite{gartner-cite}. 

Programs running on the blockchain are known as {\it smart contracts}. 
The popular Ethereum blockchain provides a low-level stack-based bytecode language that executes on top of the Ethereum Virtual Machine (EVM).
High-level languages such as Solidity and Serpent have been developed to enable traditional application developers to author smart contracts. 
However, because blockchain transactions are immutable, bugs in smart contract code have devastating consequences, and vulnerabilities in smart contracts have resulted in several high-profile exploits that undermine  trust in the underlying blockchain technology.
For example, the infamous \texttt{TheDAO} exploit~\cite{thedao-cite} resulted in the loss of almost \$60 million worth of Ether, and the \texttt{Parity Wallet} bug  caused  169 million USD worth of ether to be locked forever~\cite{parity-wallet-bug}.
The only remedy for these incidents was to hard-fork the blockchain and revert one of the forks  back to the state before the incident. 
However, this remedy itself is devastating as it defeats the core values of blockchain, such as immutability, decentralized trust, and self-governance. 
This situation leaves no  options for smart contract programmers other than writing correct code to start with.

Motivated by the serious consequences of bugs in smart contract code, recent work has studied many types of security bugs such as reentrancy, integer underflow/overflow, and issues related to delegatecalls on Ethereum. 
While these low-level bugs have drawn much attention due to high-visibility incidents on public blockchains,  we  believe that the BaaS infrastructure and enterprise scenarios bring a set of interesting, yet less well-studied security problems.

In this paper, we present our research on smart contract correctness in the context of Azure Blockchain, a BaaS solution offered by Microsoft~\cite{azure-blockchain}. 
Specifically, we focus on a  cloud service named {\it Azure Blockchain Workbench} (or Workbench for short)~\cite{workbench-cite,azure-blockchain-github}.  
The Workbench allows an enterprise customer to easily build and deploy a smart contract application  integrating active directory, database, web UI, blob storage, etc. 
A customer implements the smart contract application (that meets the requirements specified in an application policy) and uploads it onto the Workbench. 
The code is then deployed to the underlying blockchain ledger to function as an end-to-end application.
In addition to customer (application) smart contracts, the Workbench system itself is comprised of smart contracts that customize the underlying distributed blockchain consensus protocols.
Workbench ships one such smart contract for the {\it governance} of the Ethereum blockchain that uses the Proof-of-Authority (PoA) consensus protocol for validating transactions.
Workbench  relies on the correctness of the PoA governance contract to offer a trusted blockchain on Azure.

Customer contracts in the Workbench architecture implement complex business logic, starting with a high-level finite-state-machine (FSM) {\it workflow} policy specified in a JSON file. 
Intuitively, the workflow describes (a) a set of categories of users called {\it roles},  (b) the different states of a contract, and (c) the set of enabled actions (or functions) at each state restricted to each role. 
The high-level policy is useful to design contracts around state machine abstractions as well as specify the required {\it access-control} for the actions. 
While these state machines offer powerful abstraction patterns during  smart contract design, it is non-trivial to decide whether a given smart contract faithfully implements the intended FSM. 
In this paper, we define  \emph{semantic conformance checking}  as the problem of deciding whether a customer contract correctly implements the underlying workflow policy expressed as an FSM. 
Given a Workbench policy $\Policy$ that describes the workflow and a contract $\mathcal{C}$, our approach first constructs a new contract $\mathcal{C}'$ such that  $\mathcal{C}$  semantically conforms to $\Policy$ if and only if $\mathcal{C'}$ does not fail any assertions. 

In order to automatically check the correctness of the assertions in a smart contract (such as $\mathcal{C}'$ or PoA governance), we develop a new verifier called \verisol{} for smart contracts written in Solidity.  
\verisol{} is a general-purpose Solidity verifier and is not tied to Workbench. 
The verifier encodes the semantics of Solidity programs into a low-level intermediate verification language Boogie and leverages the well-engineered Boogie verification pipeline~\cite{barnett-fmco05} for both verification and counter-example generation.
In particular, \verisol{} takes advantage of existing bounded model checking tool \corral~\cite{lal-cav12} for Boogie to generate witnesses to assertion violations, and it leverages practical verification condition generators for Boogie to automate correctness proofs. 
In addition, \verisol{} uses monomial predicate abstraction~\cite{houdini, monomial} to automatically infer so-called \emph{contract invariants}, which we have found to be crucial for automatic verification of semantic conformance. 
To evaluate the effectiveness and efficiency of \verisol{}, we have performed an experiment on all $11$ sample applications that are shipped with the Workbench, as well as the PoA governance contract for the blockchain itself.
\verisol{} finds $4$ previously unknown defects in these published smart contracts, all of which have been confirmed as true bugs by the developers (many of them fixed at the time of writing).
The experimental results also demonstrate the practicality of \verisol{} in that it can perform full verification of all the fixed contracts with modest effort; most notably, \verisol{} can automatically verify $10$ out of $11$ of the fixed versions of sample smart contracts within $1.7$ seconds on average.

\vspace{5pt}
\noindent \textbf{Contributions.}
This paper makes the following contributions:
\begin{enumerate}[leftmargin=*]
\item We study the safety and security of smart contracts present in  Workbench, a BaaS offering. 
\item We formalize the Workbench application policy language and define the {\it semantic conformance} checking problem between a contract and a policy.
\item We develop a new formal verifier \verisol{} for smart contracts written in Solidity.
\item We perform an evaluation of \verisol{} on all the contracts shipped with Workbench. This includes all the application samples as well as the highly-trusted PoA governance contract.
\item We report previously unknown bugs that have been confirmed and several already fixed.
\end{enumerate}

\begin{figure}[t]
\centering
\includegraphics[scale=0.7]{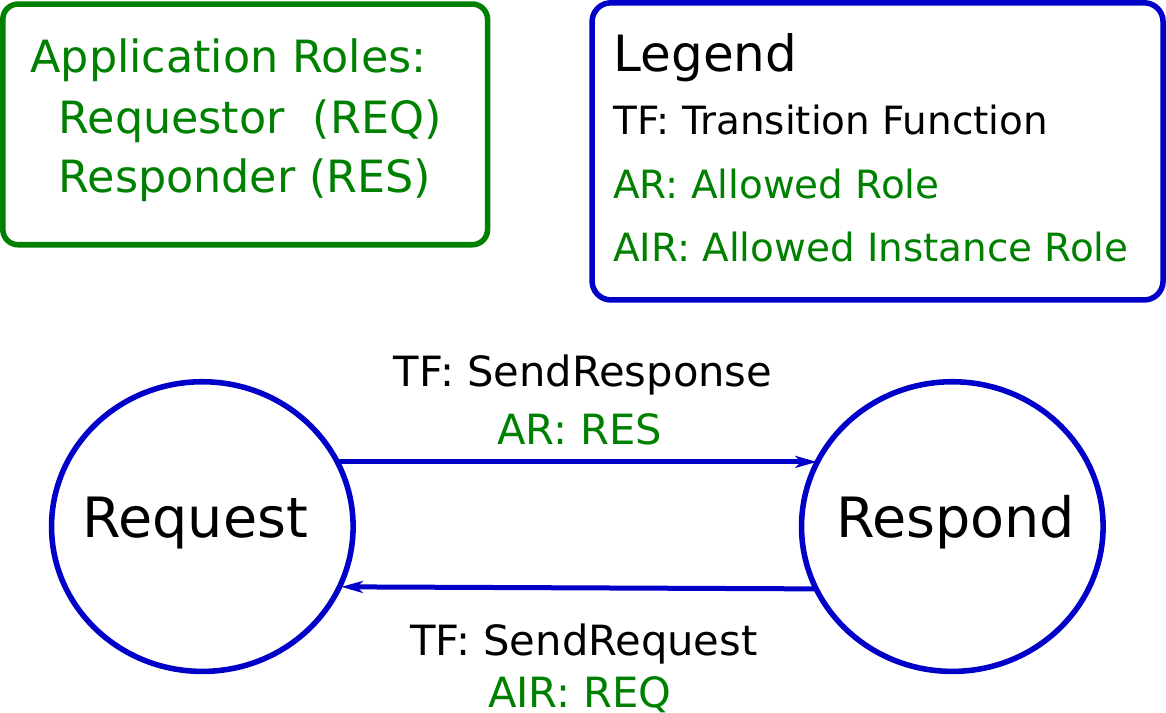}
\caption{Workflow policy diagram for HelloBlockchain application.}
\label{fig:helloblockchain-fig}
\end{figure}

\section{Overview}\label{sec:overview}

In this section, we give an example of a Workbench application policy for a sample contract and describe our approach for semantic conformance checking between the contract and the policy.

\newcommand{\helloBlockchain}{\texttt{HelloBlockchain}}

\newcommand{\requestorRole}{\textsc{Requestor}}
\newcommand{\responderRole}{\textsc{Responder}}
\newcommand{\requestState}{\texttt{Request}}
\newcommand{\respondState}{\texttt{Respond}}
\newcommand{\requestMessage}{\texttt{RequestMessage}}
\newcommand{\responseMessage}{\texttt{ResponseMessage}}
\newcommand{\instanceRequestor}{\texttt{Requestor}}
\newcommand{\instanceResponder}{\texttt{Responder}}
\newcommand{\sendRequest}{\texttt{SendRequest}}
\newcommand{\sendResponse}{\texttt{SendResponse}}

\subsection{Workbench Application Policy}
Workbench requires every customer to provide a \emph{policy} (or \emph{model}) representing the high-level workflow of the application in a JSON file\footnote{\url{https://docs.microsoft.com/en-us/azure/blockchain/workbench/configuration}}.
The policy  consists of several attributes such as the application name and description, a set of \emph{roles}, as well as  a set of \emph{workflows}. 
For example, Figure~\ref{fig:helloblockchain-fig} provides an informal pictorial representation of the policy for a simple application called \helloBlockchain.~\footnote{The example related details can be found on the associated web page: {\url{https://github.com/Azure-Samples/blockchain/tree/master/blockchain-workbench/application-and-smart-contract-samples/hello-blockchain}}}.
The application consists of two \emph{global} roles (see ``Application Roles''), namely \requestorRole{} and \responderRole{}.
Informally, each role represents a set of user addresses and provides access control or permissions for various actions exposed by the application.
We distinguish a global role from an \emph{instance role} in that the latter applies to a specific instance of the workflow. 
It is expected that the instance roles are always a subset of the user addresses associated with the global role.

As illustrated  in Figure~\ref{fig:helloblockchain-fig}, the simple \helloBlockchain{} application consists of a single workflow with two states, namely \requestState{} and \respondState{}.
The data members (or fields) include instance role members (\instanceRequestor{} and \instanceResponder{}) that range over user addresses. 
The workflow consists of two actions (or functions) in addition to the constructor function, \sendRequest{} and \sendResponse{}, both of which take a string as argument.

A transition in the workflow consists of a start state, an action or function, an access control list, and a set of successor states. 
Figure~\ref{fig:helloblockchain-fig} describes two transitions, one from each of the two states. 
For example, the application can transition from \requestState{} to \respondState{}  if a user belongs to the \responderRole{} role  ({\texttt AR})
and invokes the action \sendResponse{}. 
An ``Application Instance Role'' ({\texttt AIR}) refers to an instance role data member of the workflow that stores a member of a global role (such as \instanceRequestor{}). 
For instance, the transition  from \respondState{} to \requestState{} in Figure~\ref{fig:helloblockchain-fig} uses an \texttt{AIR} and is only allowed if the user address matches the value stored in the instance data variable \instanceRequestor{}. 

\definecolor{eminence}{RGB}{108,48,130}
\definecolor{commentgreen}{RGB}{2,112,10}
\definecolor{weborange}{RGB}{255,165,0}
\definecolor{frenchplum}{RGB}{129,20,83}
\definecolor{blue}{RGB}{0,0,130}
\definecolor{red}{RGB}{130,0,0}

\lstdefinestyle{Sol}{basicstyle=\ttfamily \scriptsize,
        language=Java,
        numbers=left,
        commentstyle=\color{commentgreen},
        numberstyle=\tiny\color{linenumbergray},
        moredelim=**[is][\btHL]{@i@}{@i@},
        moredelim=**[is][{\btHL[fill=gitdel]}]{@d@}{@d@},
        moredelim=**[is][{\btHL[fill=gitdelfocused, rounded corners=2pt]}]{@df@}{@df@},
        moredelim=**[is][{\btHL[fill=gitadd]}]{@a@}{@a@},
        moredelim=**[is][{\btHL[fill=gitaddfocused, rounded corners=2pt]}]{@af@}{@af@},
        escapeinside={(*@}{@*)},
        keywords={%
          modifier, contract,enum, function,require,address,returns,string,revert,var,int,bool,call,return,assert,assume,goto,havoc,modifies,requires,ensures,while,if, mapping,storage,memory,constructor
        },
        keywordstyle=\color{frenchplum},
        morecomment=[l]{//},
        mathescape=true,
}

\lstdefinestyle{Boogie}{basicstyle=\ttfamily \scriptsize,
        language=Java,
        numbers=left,
        numberstyle=\tiny\color{linenumbergray},
        moredelim=**[is][\btHL]{@i@}{@i@},
        moredelim=**[is][{\btHL[fill=gitdel]}]{@d@}{@d@},
        moredelim=**[is][{\btHL[fill=gitdelfocused, rounded corners=2pt]}]{@df@}{@df@},
        moredelim=**[is][{\btHL[fill=gitadd]}]{@a@}{@a@},
        moredelim=**[is][{\btHL[fill=gitaddfocused, rounded corners=2pt]}]{@af@}{@af@},
        escapeinside={(*@}{@*)},
        keywords={%
          procedure, proc, axiom, function,requires, ensures, returns, return, var,int,bool, call,assert,assume,goto,havoc,modifies,while,if,else,const,foreach
        },
        keywordstyle=\color{blue},
        commentstyle=\color{red},
        morecomment=[l]{//},
        mathescape=true,
}

\begin{figure}[htbp!]
\begin{lstlisting}[style=Sol,numbers=none]
pragma solidity ^0.4.20;

contract HelloBlockchain {
     //Set of States
    enum StateType {Request, Respond}

    //List of properties
    StateType public  State;

    address public  Requestor;
    address public  Responder;
    string public RequestMessage;
    string public ResponseMessage;
    // constructor function
    function HelloBlockchain(string message)
                             $\underline{\textrm{constructor\_checker()}}$
                             public
    {
        Requestor = msg.sender;
        RequestMessage = message;
        State = StateType.Request;
    }
    // call this function to send a request
    function SendRequest(string requestMessage) 
                         $\underline{\textrm{SendRequest\_checker()}}$ public
    {
        RequestMessage = requestMessage;
        State = StateType.Request;
    }
    // call this function to send a response
    function SendResponse(string responseMessage) 
                         $\underline{\textrm{SendResponse\_checker()}}$ public
    {
        Responder = msg.sender;
        ResponseMessage = responseMessage;
        State = StateType.Respond;
    }
    $\underline{\texttt{<modifier definitions>}}$
 }
    \end{lstlisting}
\vspace{-0.1in}
\caption{Solidity contract for HelloBlockchain application.}
\label{fig:helloblockchain-smart-contract}
\end{figure}

\subsection{Workbench Application Smart Contract}
After specifying the application policy, a user provides a smart contract for the appropriate blockchain ledger to implement the workflow.
Currently, Workbench supports the popular language Solidity for targeting applications on Ethereum. 
Figure~\ref{fig:helloblockchain-smart-contract} describes a Solidity smart contract that implements the \helloBlockchain{} workflow in the \helloBlockchain{} application.
For the purpose of this sub-section, we start by ignoring the portions of the code that are $\underline{\texttt{underlined}}$.
The contract declares the data members present in the configuration as state variables with suitable types. 
Each contract implementing a workflow defines an additional state variable  \texttt{State} to track the current state of a workflow. 
The contract consists of the constructor function along with the two functions defined in the policy, with matching signatures. 
The functions set the state variables and update the state variables appropriately to reflect the state transitions. 

The Workbench service allows a user to upload the policy, the Solidity code, and optionally add users and perform various actions permitted by the configuration. 
Although the smart contract drives the application, the policy is used to expose the set of enabled actions at each state for a given user. 
Discrepancies between the policy and Solidity code can lead to unexpected state transitions that do not conform to the high-level policy. To ensure the correct functioning and security of the application, it is crucial to verify that the Solidity program semantically conforms to the intended meaning of the policy configuration.

\subsection{Semantic Conformance Verification}

Given the application policy and a smart contract, we define the problem of {\it semantic conformance} between the two that ensures that the smart contract respects the policy (Section~\ref{sec:semantic-conformance}).
Moreover, we reduce the semantic conformance verification problem to checking assertions on an instrumented Solidity program. 
For the \helloBlockchain{} application, the instrumentation is provided by adding the \underline{underlined} {\it modifier} invocations in Figure~\ref{fig:helloblockchain-smart-contract}.
A {\it modifier} is a Solidity language construct that allows wrapping a function invocation with code that executes before and after the execution.

\newcommand{\SolidityNonDetFn}{\texttt{nondet}}

\begin{figure}[htbp]
\begin{lstlisting}[style=Sol,numbers=none]
   function nondet() returns (bool); //no definition  

    // Checker modifiers 
    modifier constructor_checker() 
    {
       _;      
      assert (nondet() /* global role REQUESTOR */
                  ==> State == StateType.Request);
    }
    modifier SendRequest_checker() 
    {
      StateType oldState = State;
      address oldRequestor = Requestor;
       _;
      assert ((msg.sender == oldRequestor && 
               oldState == StateType.Respond) 
              ==> State == StateType.Request);
    }
    modifier SendResponse_checker() 
    {
      StateType oldState = State;
       _;
      assert ((nondet() /* global role RESPONDER */ &&
               oldState == StateType.Request) 
              ==> State == StateType.Respond);
    }

\end{lstlisting}
\vspace{-0.1in}
\caption{Modifier definitions for instrumented HelloBlockchain application.}
\label{fig:instrumented-running-example}
\end{figure}

Figure~\ref{fig:instrumented-running-example} shows the definition of the modifiers used to instrument for conformance checking. 
Intuitively, we wrap the constructor and functions with checks to ensure that they implement the FSM state transitions correctly.
For example, if the FSM transitions from a state $s_1$ to a state $s_2$ upon the invocation of function $f$ by a user with access control $\mathit{ac}$, then we instrument the definition of $f$ to ensure that any execution starting in $s_1$ with access control satisfying $\mathit{ac}$ should transition to $s_2$.

Finally, given the instrumented Solidity program, we discharge the assertions statically using a new formal verifier for Solidity called \verisol{}.
The verifier can find counterexamples (in the form of a sequence of transactions involving calls to the constructor and public methods) as well as automatically construct proofs of semantic conformance. 
Note that, even though the simple \varn{HelloBlockchain} example does not contain any unbounded loops or recursion, verifying semantic conformance still requires reasoning about executions that involve unbounded numbers of calls to the two public functions. 
We demonstrate that \verisol{} is able to find deep violations of the conformance property for well-tested Workbench applications, as well as automatically construct inductive proofs for most of the application samples shipped with Workbench.


\section{Semantic Conformance Checking for Workbench Policies}
\label{sec:conformance}

\newcommand{\workbenchInterface}{{\em WBAppPolicy}}
\newcommand{\WfAltFunction}[1]{\mathit{g}_{#1}}

In this section, we formalize the Workbench application policy that we informally introduced in Section~\ref{sec:overview}. 
Our formalization can be seen as a mathematical representation of the official Workbench application JSON schema documentation.

\newcommand{\Appl}{\it{App}}
\newcommand{\Roles}[1]{\mathcal{R}_{#1}}
\newcommand{\role}[1]{\mathit{r}_{#1}}
\newcommand{\Workflows}[1]{\mathcal{W}_{#1}}
\newcommand{\workflow}[1]{\mathit{w}_{#1}}
\newcommand{\WfStates}[1]{\mathcal{S}_{#1}}
\newcommand{\WfState}[1]{\mathit{s}_{#1}}
\newcommand{\WfStateVar}{\mathit{State}}
\newcommand{\WfProperties}[1]{\mathcal{P}_{#1}}
\newcommand{\WfInstanceProperties}[1]{\mathcal{R}_{w#1}}
\newcommand{\WfInstanceProperty}[1]{\mathit{q}_{#1}}
\newcommand{\WfProperty}[1]{\mathit{p}_{#1}}
\newcommand{\WfInitState}[1]{\WfState{0#1}}
\newcommand{\WfFunctions}[1]{\mathcal{F}_{#1}}
\newcommand{\WfFunction}[1]{\mathit{f}_{#1}}
\newcommand{\Types}[1]{\mathcal{T}_{#1}}
\newcommand{\type}[1]{\mathit{t}_{#1}}
\newcommand{\param}[1]{\mathit{x}_{#1}}
\newcommand{\WfTransitions}[1]{\mathcal{\gamma}_{#1}}
\newcommand{\Identifier}[1]{\mathit{id}_{#1}}
\newcommand{\AccessControl}[1]{\mathcal{AC}_{#1}}
\newcommand{\accessSet}[2]{\mathit{ac}_{#2}^{#1}}
\newcommand{\WfRoleRegistry}{\mathtt{RoleRegistry}}
\newcommand{\sender}{\mathit{sender}}

\newcommand{\translateAC}[1]{P(#1)}
\newcommand{\translatePostStates}[1]{\alpha(#1)}

\newcommand{\intType}{\mathit{integer}}
\newcommand{\stringType}{\mathit{string}}
\newcommand{\addrType}{\mathit{address}}

\subsection{Formalization of Workbench Application Policies}
\label{sec:workbench-interface}

The Workbench policy for an application allows the user to describe (i) the {\it data members} and {\it actions} of an application, (ii) a high-level {\it state-machine} view of the application, and (iii) role-based {\it access control} for {\it state transitions}.
The role-based access control provides security for deploying smart contracts in an open and adversarial setting; 
the high-level state machine naturally captures the essence of a {\it workflow} that progresses between a set of states based on some actions from the user.


More formally, a {\it Workbench Application Policy} is a pair $(\Roles{}, \Workflows{})$ where $\Roles{}$ is a set of   {\it global roles} used for access control, and $\Workflows{}$ is a set of {\it workflows} defining a kind of finite state machine. Specifically, a workflow is defined  by a tuple $\langle \WfStates{}, \WfInitState{},   \WfInstanceProperties{}, \WfFunctions{}, \WfFunctions{0}, \accessSet{}{0}, \WfTransitions{}\rangle$ where:

\begin{itemize}[leftmargin=*] 
\item $\WfStates{}$ is a finite set of \emph{states}, and $\WfInitState{} \in \WfStates{}$ is an \emph{initial state}
\item $\WfInstanceProperties{}$  is a finite set of \emph{instance roles} of the  form $(\Identifier{}: \type{})$, where $\Identifier{}$ is an identifier and $\type{}$ is a role drawn from $\Roles{}$

\item $\WfFunctions{}(\Identifier{0}, \ldots, \Identifier{k})$ is a set of {\it actions (functions)}, with $\WfFunctions{0}$ denoting an initial action (constructor)
\item  $\accessSet{}{0} \subseteq \Roles{}$  is the \emph{initiator role}  for restricting users that can create an instance of the contract 
\item  $\WfTransitions{} \subseteq \WfStates{} \times \WfFunctions{} \times (\WfInstanceProperties{} \cup \Roles{}) \times 2^{\WfStates{}}$ is a set of transitions. Given a transition $\tau = (s, f, ac, S)$, we write $\tau.s, \tau.f, \tau.ac, \tau.S$ to denote $s, f, ac$, and $S$ respectively
\end{itemize} 
Intuitively, $\WfStates{}$ defines the different ``states'' that the contract can be in, and  $\WfTransitions{}$ describes which state can transition to what other states by performing certain actions. The transitions are additionally ``guarded'' by  roles (which can be either global or instance roles) that qualify which users are allowed to perform those actions. As mentioned earlier in Section~\ref{sec:overview}, each ``role'' corresponds to a set of users (i.e., addresses on the blockchain). The use of instance roles in the workbench policy allows different instances of the contract to authorize different users to perform certain actions.

\subsection{Semantic Conformance}
\label{sec:semantic-conformance}

Given a  contract $\Contract$ and a Workbench Application policy $\Policy$, \emph{semantic conformance} between $\Contract$ and $\Policy$ requires that the contract $\Contract$ faithfully implements the policy specified by $\Policy$. 
In this subsection, we first  define some syntactic requirements on the contract, and then formalize what we mean by semantic conformance between a contract and a policy.

\vspace{5pt}
\noindent \textit{Syntactic conformance.} Given a client contract $\Contract$ and a policy $\Policy = (\Roles{}, \Workflows{})$, our syntactic conformance requirement stipulates that the contract for each $w \in \Workflows{}$ implements all the instance state variables as well as definitions for each of the functions.  
Additionally, each contract function has a parameter called $\sender$, which is a blockchain address that denotes the user or contract invoking this function. 
Finally, each contract should contain a state variable $s_w$  that ranges over $\WfStates{w}$, for each $w \in \Workflows{}$. 

\vspace{5pt}
\noindent \textit{Semantic conformance.} We formalize  the semantic conformance requirement for smart contracts using Floyd-Hoare triples of the form $\{\phi\} \ S \ \{\psi\}$ indicating that any execution of statement $S$ starting in a state satisfying $\phi$ results in a state satisfying $\psi$ (if the execution of $S$ terminates). We can define semantic conformance between a contract $\Contract$ and a policy $\Policy$ as a set of Hoare triples, one for each pair $(m, s)$ where $m$ is a method in the contract and $s$ is a state in the Workbench policy. At a high-level, the idea is simple: we insist that, when a function  is executed along a transition, the resulting state transition should be in accordance with the Workbench policy.  

Given an application policy $\Policy =  (\Roles{}, \Workflows{})$ and workflow $w =  \langle \WfStates{}, \WfInitState{},   \WfInstanceProperties{}, \WfFunctions{}, \WfFunctions{0}, \accessSet{}{0}, \WfTransitions{}\rangle \in \Workflows{}$, we can formalize this high-level idea by using the following Hoare triples:

\newcommand{\isUserAddress}{{\mathit {isUser}}}

\begin{enumerate}[leftmargin=*]
\item {\bf Initiation}. 
\[
\begin{array}{c}
\{ \sender \in \accessSet{}{0}\}  \ \ 
 \mathcal{F}_0(v_1, \ldots, v_{k}) \ \ 
\{s_w = \WfInitState{}\} 
\end{array}
\]
The Hoare triple states that the creation of an instance of the workflow with the appropriate access control $\accessSet{}{0}$ results in establishing the initial state.



\item {\bf Consecution}. Let  $\tau = (\WfState{1}, \WfFunction{}, \accessSet{}{}, \WfStates{2})$ be a transition in  $\WfTransitions{}$. Then, for each such transition, semantic conformance requires the following Hoare triple to be valid:
\[
\begin{array}{c}
\{\sender \in \accessSet{}{} \wedge s_w = \WfState{1}\}   \ \ 
 \WfFunction{}(v_1, \ldots, v_{k})   \ \ 
\{ s_w \in \WfStates{2}\}
\end{array}
\]

Here, the precondition checks two facts: 
First, the $\sender$ must satisfy the access control, and, second, the start state must be $\WfState{1}$. The  post-condition asserts that the implementation of method $f$ in the contract results in a state that is valid according to policy $\Policy$.
\end{enumerate}

\subsection{Instrumentation for Semantic Conformance Checking}
\label{sec:solidity-instrumentation}
As mentioned in Section~\ref{sec:overview}, our approach checks semantic conformance of Solidity contracts by (a) instrumenting the contract with assertions, and (b) using a verification tool to check that none of the assertions can fail. We explain our instrumentation strategy in this subsection and refer the reader to Section~\ref{sec:verification} for a description of our verification tool chain.

\newcommand{\solidityExpression}[1]{\parallel{#1}\parallel}

Our instrumentation methodology heavily relies on the  \texttt{modifier} construct in Solidity.
A modifier has syntax very similar to a function definition in Solidity with a name and list of parameters and a body that can refer to parameters and globals in scope.
The general structure of a modifier definition without any parameters  is~\cite{solidity-modifiers}:
\begin{lstlisting}[style=Sol,numbers=none,basicstyle=\ttfamily  \small]
modifier Foo() {   
   pre-stmt;
   _; 
   post-stmt;
}
\end{lstlisting}
where \texttt{pre-stmt} and \texttt{post-stmt} are Solidity statements. 
When this modifier is applied to a function \texttt{Bar}, 
\begin{lstlisting}[style=Sol,numbers=none,basicstyle=\ttfamily  \small]
function Bar(int x) Foo(){   
   Bar-stmt;
}
\end{lstlisting}
the Solidity compiler transforms the body of \texttt{Bar} to execute \texttt{pre-stmt} (respectively, \texttt{post-stmt}) before (respectively, after) \texttt{Bar-stmt}.
This provides a convenient way to inject code at multiple return sites from a procedure and also inject code before the execution of the constructor code (since a constructor may invoke other base class constructors implicitly). 

We now define a couple of helper predicates before describing the actual checks. 
Let $\translateAC{\accessSet{}{}}$ be a predicate that encodes the membership of $\sender$ in the set $\accessSet{}{}$:

\[
\small{
\begin{array}{lll}
\translateAC{\accessSet{}{}} & \doteq & \left\{
  \begin{array}{ll}
  \mathit{false}, & \accessSet{}{} = \{\} \\
  \mathtt{msg.sender} = q, & \accessSet{}{} = \{q \in \WfInstanceProperties{}\} \\
  \mathtt{\SolidityNonDetFn()} & \accessSet{}{} = \{r \in \Roles{}\} \\
  \translateAC{\accessSet{1}{}} \vee \translateAC{\accessSet{2}{}} & \accessSet{}{} = \accessSet{1}{} \cup \accessSet{2}{}
  \end{array}
  \right.
\end{array}
}
\]
Here $\SolidityNonDetFn$ is a side-effect free Solidity function that returns a non-deterministic Boolean value at each invocation.
For the sake of static verification, one can declare a function without any definition. 
This allows us to model the membership check $\sender \in \accessSet{}{}$ conservatively in the absence of global roles on the blockchain.

Next, we define a predicate for membership of a contract state in a set of states $\WfStates{}' \subseteq \WfStates{}$ using $\translatePostStates{\WfStates{}'}$ as follows:
\[
\small{
\begin{array}{lll}
\translatePostStates{\WfStates{}'} & \doteq & \left\{
  \begin{array}{ll}
  \mathit{false}, & \WfStates{}' = \{\} \\
  s_w = s, &  \WfStates{}' = \{\WfState{} \in \WfStates{}\} \\
  \translatePostStates{\WfStates{1}} \vee \translatePostStates{\WfStates{2}}, & \WfStates{}' = \WfStates{1} \cup \WfStates{2} \\
  \end{array}
  \right.
\end{array}
}
\]

We can now use these predicates to define the source code transformations below:

\vspace{5pt}
\noindent \textit{Constructor.}  We add the following modifier to constructors:

\begin{lstlisting}[style=Sol,numbers=none,basicstyle=\ttfamily  \small]
modifier constructor_checker() {   
   _;
   assert ($\translateAC{\accessSet{}{0}} \Rightarrow \translatePostStates{\{\WfInitState{}\}}$);
}
\end{lstlisting}
Here, the assertion ensures that the constructor sets up the correct initial state when executed by a user with access control $\accessSet{}{0}$.

\vspace{5pt}
\noindent \textit{Other functions.} 
For a function $\WfAltFunction{}$, let  $\WfTransitions{}^{\WfAltFunction{}} \doteq \{\tau \in \WfTransitions{} \ | \ \tau =  (\WfState{1}, g{}, \accessSet{}{}, \WfStates{2}) \}$ be the set of all transitions where $\WfAltFunction{}$ is invoked. 

\begin{lstlisting}[style=Sol,numbers=none, basicstyle=\ttfamily \small]
modifier g_checker() {
   // copy old State
   StateType oldState = $s_w$;
   // copy old instance role vars
   ...
   _;
   assert $\bigwedge_{\tau \in \WfTransitions{}^{\WfAltFunction{}}} \left(\texttt{old}\left(\translateAC{\tau.\accessSet{}{}}  \wedge \translatePostStates{\{\tau.\WfState{}\}}\right) \Rightarrow \translatePostStates{\tau.\WfStates{}}\right)$;
}
\end{lstlisting}



Here, the instrumented code first copies the $s_w$ variable and all of the variables in $\WfInstanceProperties{}$ into corresponding ``old'' copies.
Next, the assertion checks that if the function is executed in a transition $\tau$, then state transitions to one of the successor states  in $\tau.\WfStates{}$.
The notation $\texttt{old}(e)$ replaces any occurrences of a state variable (such as $s_w$) with the ``old'' copy that holds the value at the entry to the function.
Figure~\ref{fig:instrumented-running-example} shows the modifier definitions for our running example \helloBlockchain{} described in Section~\ref{sec:overview}
Although we show the $\SolidityNonDetFn()$ to highlight the issue of global roles, one can safely replace $\SolidityNonDetFn()$ with $\textit{true}$ since the function only appears negatively in any assertion.
In fact, this observation allows us to add the simplified assertions for runtime checking as well. 
Finally, since conjunction distributes over assertions, we can replace the single assertion with an assertion for each transition in the implementation. 

\newcommand{\irule}[2]%
    {\mkern-2mu\displaystyle\frac{#1}{\vphantom{,}#2}\mkern-2mu}
\newcommand{\irulelabel}[3]
{
\mkern-2mu
\begin{array}{ll}
\displaystyle\frac{#1}{\vphantom{,}#2} & \!\!\!\!#3
\end{array}
\mkern-2mu
}

\newcommand{\hoare}[3]{\{#1\}~#2~\{#3\}}
\newcommand{\inv}{\mathcal{I}}

\newcommand{\this}{\texttt{this}}
\newcommand{\length}{\texttt{Length}}
\newcommand{\dtype}{\tau}
\newcommand{\alloc}{\texttt{Alloc}}
\newcommand{\malloc}{\textrm{New}}
\newcommand{\push}{\texttt{push}}
\newcommand{\msgsender}{\varn{msg\_sender}}

\section{Formal Verification using \verisol{}}
\label{sec:verification}

In this section, we present our formal verifier called \verisol{} for checking the correctness of assertions in Solidity smart contracts. 
Since our verifier is built on top of the Boogie tool chain, it can be used for both verification and counterexample generation.

\subsection{General Methodology} \label{sec:methodology}
Let $\Contract = \set{\lambda \vec{x_0}.f_0,~ \lambda \vec{x_1}.f_1,~ \ldots,~ \lambda \vec{x_n}.f_n }$ be a smart contract  annotated with assertions where:
\begin{itemize}
  \item $\lambda \vec{x_0}.f_0$ is the constructor
  \item $\lambda \vec{x_i}.f_i$ for $i \in [1, n]$ are public functions
\end{itemize}
Our verification methodology is based on finding a \emph{contract invariant} $\inv$ satisfying the following Hoare triples:
\begin{equation} \label{eq:hoare1}
  \models \hoare{\texttt{true}}{f_0}{\inv}
\end{equation}
\begin{equation} \label{eq:hoare2}
  \models \hoare{\inv}{f_i}{\inv} \text{~for all~} i \in [1, n]
\end{equation}

\begin{figure}[t]
\begin{lstlisting}[style=Boogie, numbers=none, xleftmargin=5pt, numbersep=5pt, basicstyle=\ttfamily \small]
call $f_0$(*);
while (true) {
    if (*) call $f_1$(*);
    else if (*) call $f_2$(*);
    ...
    else if (*) call $f_n$(*);
}
\end{lstlisting}
\caption{Harness for Solidity contracts}
\label{fig:harness}
\end{figure}

Here, the first condition states the contract invariant is established by the constructor, and the second condition states that $\inv$ is inductive --- i.e., it is preserved by every public function in $\Contract$. Note that such a contract invariant suffices to establish the validity of all assertions in the contract under \emph{any} possible sequence of function invocations of the contract. To see why this is the case, consider a ``harness'' that invokes the functions in $\Contract$ as in Figure~\ref{fig:harness}.
This harness first creates an instance of the contract by calling the constructor, and then repeatedly and non-deterministically invokes one of the public functions of $\Contract$. Observe that the Hoare triples (1) and (2) listed above essentially state that $\inv$ is an inductive invariant of the loop in this harness; thus, the contract invariant $\inv$ overapproximates the state of the contract under any sequence of the contract's function invocations. Furthermore, when the functions contain assertions, the Hoare triple $\{\inv\} \ f_i \ \{\inv\}$ can only be proven if $\inv$ is strong enough to imply the assertion conditions. Thus, the validity of the Hoare triples in (1) and (2) establishes  correctness under all possible usage patterns of the contract.


\subsection{Overview}
\label{sec:verif-overview}

\begin{figure}[t]
\centering
\includegraphics[scale=0.55]{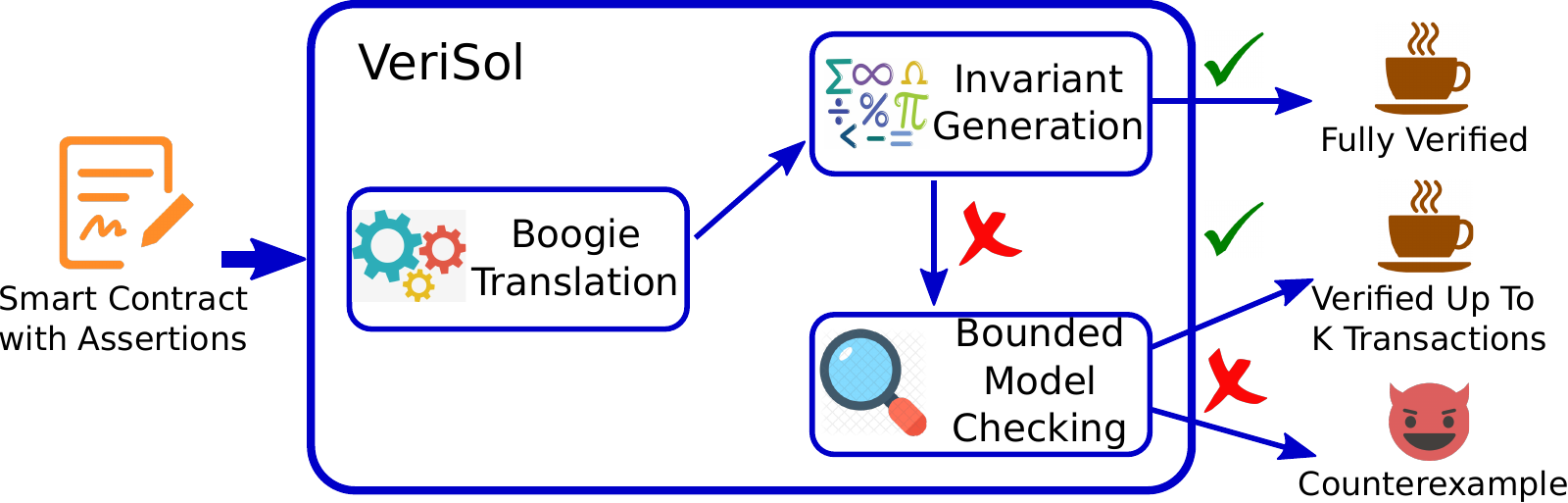}
\caption{Schematic workflow of \verisol{}.}
\label{fig:workflow}
\end{figure}

We now describe the design of our tool called \verisol{} for checking safety of smart contracts. \verisol{} is based on the proof methodology outlined in Section~\ref{sec:methodology}, and its workflow is illustrated in Figure~\ref{fig:workflow}. At a high-level, \verisol{} takes as input a Solidity contract $\Contract$ annotated with assertions and  yields one of the following three outcomes:
\begin{itemize}[leftmargin=*]
\item \emph{Fully verified:} This means that the assertions in $\Contract$ are guaranteed not to fail under any usage scenario.
\item \emph{Refuted:} This  indicates that $\Contract$ was able to find at least one input and invocation sequence of the contract functions under which one of the assertions is guaranteed to fail.
\item  \emph{Partially verified:}  When \verisol{} can neither verify nor refute contract correctness, it performs bounded verification to establish that the contract is safe up to $k$ transactions. This essentially corresponds to unrolling the ''harness'' loop from Figure~\ref{fig:harness} $k$ times and then verifying that the assertions do not fail in the unrolled version. 
\end{itemize} 

\verisol{} consists of three modules, namely (a) {\it Boogie Translation} from a Solidity program, (b) {\it Invariant Generation} to infer a {\it contract invariant} as well as loop invariants and procedure summaries, and (c) {\it Bounded Model Checking} to explore assertion failures within all transactions up to a user-specified depth $k$.
In the remaining subsections, we discuss each of these components in more detail.

\subsection{Solidity to Boogie Translation} 

In this subsection, we  formally describe our translation of Solidity source code to the Boogie intermediate verification language.  We start with a brief description of Solidity and Boogie, and then discuss our translation.

\newcommand{\solTypes}{\mathit{SolTypes}}
\newcommand{\solType}{\mathit{st}}
\newcommand{\solBaseTypes}{\mathit{SolElemTypes}}
\newcommand{\solBaseType}{\mathit{et}}
\newcommand{\solContractTypes}{\mathit{ContractNames}}
\newcommand{\solContract}{\mathit{ct}}
\begin{figure}[!t]
\[
\begin{array}{lll}
\solContract \in \solContractTypes \\
\solBaseType \in \solBaseTypes & ::= & \intType \ | \ \stringType \ | \ \addrType \\
\solType \in \solTypes & ::= & \solBaseType \ | \ \solContract \ | \ \solBaseType \Rightarrow \solType
\end{array}
\]
\[
\begin{array}{lcl}
\solExpr{} \in \solExprs & \!\!\!::=\!\!\! & c \ | \  \varn{x}  \ | \  \bop{\solExpr{},\ldots, \solExpr{}}  \ | \ \arrSel{\solExpr{}}{\solExpr{}} \\
                    & | & \varn{msg.sender} \ | \ \varn{x.length} \\
\solStmt{} \in \solStmts & \!\!\!::=\!\!\! & \bassign{\varn{x}}{\solExpr{}} \ | \ \varn{x}[\solExpr{}]\ldots[\solExpr{}] := \varn{y} \ | \ \Seq{\solStmt{}}{\solStmt{}} \\
                    & | & \solRequire(\solExpr{}) \ | \ \solAssert(\solExpr{}) \\
                    & | & \solIfelse{\solExpr{}}{\solStmt{}}{\solStmt{}} \\
                    & | & \solWhile{\solExpr{}}{\solStmt{}} \ | \ \varn{x}.\push(\solExpr{}) \\
                    & | & \solExpr{} := \varn{f}(\vec{\solExpr{}}) \ | \ \solExpr{} := \solExpr{}.\varn{f}(\vec{\solExpr{}}) \\
                    & | & \solExpr{} := \texttt{new} \ C(\vec{\solExpr{}}) \ | \ \solExpr{} := \texttt{new} \ t[](\solExpr{}) \\
                    & | & \solExpr{} := \texttt{new} \ (t_1 \Rightarrow t_2 \Rightarrow \ldots \Rightarrow t_k \Rightarrow t)() \\
\end{array}
\]
\caption{A subset of Solidity language. $C$ denotes a contract name, $t$ denotes an elementary Solidity type, and $\varn{f}$ denotes a function name.}
\label{fig:solidity-lang}
\end{figure}

\vspace{5pt}
\noindent \textbf{Solidity Language.}
Figure~\ref{fig:solidity-lang} shows a core subset of Solidity that we use for our formalization.  
At a high level, Solidity is a typed object-oriented programming language with built-in support for basic verification constructs, such as the \texttt{require} construct for expressing pre-conditions. 

Types in our core language include integers, strings, contracts, addresses, and mappings. We use the notation $\tau_1 \Rightarrow \tau_2$ to denote a mapping from a value of type $\tau_1$ to a value of type $\tau_2$ (where $\tau_2$ can be a nested map type).  Since arrays can be viewed as a special form of mapping $\intType \Rightarrow \tau$, we do not introduce a separate array type to simplify presentation.

As standard, expressions in Solidity include constants, local variables, \emph{state variables} (i.e., fields in standard object-oriented language terminology), unary/binary operators (denoted $\bfont{op}$), and array/map lookup $e[e']$. Given an array \varn{x}, the expression \varn{x.length} yields the length of that array, and \varn{msg.sender} yields the address of the contract or user that initiates the current function invocation.

Statements in our core Solidity language include  assignments, conditional statements, loops, sequential composition,  array insertion (\varn{push}), internal and external function calls, contract instance creation, and dynamic  allocations. 
The construct $\solRequire$ is used to specify the precondition of a function, and $\solAssert$ checks that its input evaluates to true and terminates execution otherwise. 
Solidity differentiates between two types of function calls: internal and external. An \emph{internal} call $\solExpr{} := \varn{f}(\vec{\solExpr{}})$ invokes the function $\varn{f}$ and keeps \varn{msg.sender}  unchanged. 
An \emph{external} call $\solExpr{} := \solExpr{0}.\varn{f}(\vec{\solExpr{}})$ invokes function $\varn{f}$ in the contract instance pointed by $\solExpr{0}$ (which may include $\this$), and  uses $\this$ as the \varn{msg.sender} for the callee.

We omit several aspects of the language that are desugared into our core Solidity language. 
This includes fairly comprehensive support for {\it modifiers}, {\it libraries} and {\it structs}.

\newcommand{\bBaseTypes}{\mathit{BoogieElemTypes}}
\newcommand{\bBaseType}{\mathit{bbt}}
\newcommand{\bTypes}{\mathit{BoogieTypes}}
\newcommand{\bType}{\mathit{bt}}
\begin{figure}[!t]
\[
\begin{array}{lll}
\bBaseType \in \bBaseTypes & ::= & \ \bint  \ | \ \bRef \\
\bType \in \bTypes & ::= & \ \bBaseType \ | \ [\bBaseType]\bType
\end{array}
\]
\[
\begin{array}{lcl}
\bexpr \in \Exprs &\!\!\!::=\!\!\! & c \ | \  \varn{x}  \ | \  \bop{\bexpr,\ldots, \bexpr}  \ | \ \varn{uf}(\bexpr, \ldots, \bexpr) \\
& | &   \varn{x}[e]\ldots[e] \ | \ \forall i: \bBaseType :: \bexpr\\
\sstmt \in \Stmts &\!\!\!::=\!\!\! & \bskip \ | \ \bhavoc{\varn{x}} \ | \ \bassign{\varn{x}}{\bexpr} \\
& | &   \varn{x}[e]\ldots[e] := e \\
                  & | & \bassume{\bexpr} \ | \ \bassert{\bexpr} \\
                  & | & \bcall{\pproc}{\bexpr, \ldots, \bexpr}{\varnVector{x}} \ | \ \Seq{\sstmt}{\sstmt} \\
                  & | & \bifelse{\bexpr}{\sstmt}{\sstmt} \ | \ \bwhile{\bexpr}{\sstmt}
\end{array}
\]
\caption{A subset of Boogie language.}
\label{fig:boogie-lang}
\end{figure}

\vspace{5pt}
\noindent \textbf{Boogie Language.} Since our goal is to translate Solidity to Boogie, we also give a brief overview of the Boogie intermediate verification language.
As shown in Figure~\ref{fig:boogie-lang}, types in Boogie include integers ($\bint$), references ($\bRef$), and nested maps.
Expressions ($\Exprs$) consist of constants, variables, arithmetic and logical operators ($\bfont{op}$), uninterpreted functions (\varn{uf}), map lookups, and quantified expressions.
Standard statements ($\Stmts$) in Boogie consist of skip, variable and array/map assignment, sequential composition, conditional statements, and loops.
The $\bhavoc{\varn{x}}$ statement assigns an arbitrary value of appropriate type to a variable $\varn{x}$.
A procedure call ($\bcall{\pproc}{\bexpr, \ldots, \bexpr}{\varnVector{x}}$)  returns a vector of values that can be stored in local variables. 
The $\bassertn$ and $\bassumen$ statements behave as  no-ops  when their arguments evaluate to true and terminate execution otherwise. 
An assertion failure is considered a failing execution, whereas an assumption failure blocks.

\vspace{5pt}
\noindent \textbf{From Solidity to Boogie types.}
We define a function $\mu: \solTypes \rightarrow \bTypes$ that translates a Solidity type to a type in Boogie as follows:
\[
\small
\begin{array}{lll}
\mu({\solType}) & \doteq & \left\{
  \begin{array}{ll}
  \bint, & \solType \in \{\intType, \stringType\} \\
  \bRef, & \solType \in \{\addrType\} \cup \solContractTypes \\
  \bRef, & \solType \doteq \solBaseType \Rightarrow \solType \\
  \end{array}
  \right.
\end{array}
\]
Specifically, we translate Solidity integers and strings to Boogie integers; addresses, contract names, and mappings to Boogie references.
Note that we represent Solidity strings as integers in Boogie because Solidity only allows equality checks between strings in the core language.

\vspace{5pt}
\noindent \textbf{From Solidity to Boogie expressions.}
We present our translation from Solidity to Boogie expressions using  judgments of the form $\vdash e \hookrightarrow \chi$ in Figure~\ref{fig:enc-expr}, where $e$ is a Solidity expression and $\chi$ is the corresponding Boogie expression. 
While Solidity local variables and the expression \varn{msg.sender} are mapped directly into Boogie local variables and parameters respectively, state variables in Solidity are translated into array lookups. 
Specifically, for each  state variable \varn{x} for contract $C$, we introduce  a mapping $\varn{x}^C$ from contract instances $o$ to the value stored in its state variable \varn{x}. Thus, reads from state variable \varn{x} are modeled as $\varn{x}^C[\varn{this}]$ in Boogie. Next, we translate string constants in Solidity to Boogie integers using an uninterpreted function called \emph{StrToInt} that is applied to a hash of the string~\footnote{We assume that the hash function is collision-free. In our implementation, we enforce this by keeping a mapping from each string constant to a counter.}.  As mentioned earlier, this string-to-integer translation does not cause imprecision because Solidity only allows equality checks between variables of type string.

\begin{figure}[!t]
\centering
\[
\begin{array}{c}
\irulelabel
{\varn{x} \in \emph{LocalVars}}
{\vdash \varn{x} \hookrightarrow \varn{x}}
{\textrm{(Var1)}}
~
\irulelabel
{v = \msgsender}
{\vdash \varn{msg.sender} \hookrightarrow v}
{\textrm{(Sender)}} \\ \ \\

\irulelabel
{\emph{Type}(c) \neq \stringType}
{\vdash c \hookrightarrow c}
{\textrm{(Const1)}}
~
\irulelabel
{\varn{x} \in \emph{StateVars(C)}}
{\vdash \varn{x} \hookrightarrow \varn{x}^C[\this]}
{\textrm{(Var2)}} \\ \ \\

\irulelabel
{\begin{array}{c}
    \emph{Type}(c) = \stringType \quad c' = \emph{Hash}(c) \\
\end{array}}
{\vdash c \hookrightarrow \emph{StrToInt}(c')}
{\textrm{(Const2)}} \\ \ \\

\irulelabel
{\vdash \varn{x} \hookrightarrow \chi}
{\vdash \varn{x}.\textrm{length} \hookrightarrow \length[\chi]}
{\textrm{(Len)}} \\ \ \\


\irulelabel
{\begin{array}{c}
    \vdash e_1 \hookrightarrow \chi_1 \quad \vdash e_2 \hookrightarrow \chi_2 \\
    \emph{Type}(e_2) = t_1 \quad \emph{Type}(e_1[e_2]) = t_2 \\
\end{array}}
{\vdash e_1[e_2] \hookrightarrow M_{\mu(t_1)}^{\mu(t_2)}[\chi_1][\chi_2]}
{\textrm{(Map)}} \\ \ \\

\irulelabel
{\vdash e_i \hookrightarrow \chi_i \quad i = 1, \ldots, n}
{\vdash \bop{e_1, \ldots, e_n} \hookrightarrow \bop{\chi_1, \ldots, \chi_n}}
{\textrm{(Op)}} \\ \ \\

\end{array}
\]
\vspace{-20pt}
\caption{Inference rules for encoding Solidity expressions to Boogie expressions. $\emph{Type}(e)$ is a function that returns the static type of Solidity expression $e$.}
\label{fig:enc-expr}
\end{figure}

Similar to our handling of state variables, our Boogie encoding also introduces an array  called \varn{Length} to map each Solidity array to its corresponding length. Thus, a Solidity expression $\varn{x}.\texttt{length}$ is translated as $\varn{Length}[\chi]$ where $\chi$ is the Boogie encoding of \varn{x}.

The translation of array/map look-up is somewhat more involved due to potential aliasing issues. 
First, the basic idea is that for each map of type $t_1 \Rightarrow t_2$, we introduce a Boogie map  $M_{\tau}^{\tau'}$ where $\tau$ is the Boogie encoding of type  $t_1$ (i.e., $\tau = \mu(t_1)$) and $\tau'$ is the Boogie encoding of type $t_2$ (i.e., $\tau' = \mu(t_2)$). 
Intuitively,  $M_{\tau}^{\tau'}$ maps each array/map object to its contents, which are in turn represented as a map. 
Thus, we can think of  $M_{\tau}^{\tau'}$ as a two-dimensional mapping where the first dimension is the address of the Solidity map and the second dimension is the look-up key. 
For a nested map expression $e_1$ of type  $t_1 \Rightarrow t_2$ where $t_2$ is a nested map/array, observe that we look up $e_1$ in $M_{\mu(t_1)}^\bRef$ since maps and arrays in Solidity are dynamically allocated. 
Intuitively, everything that can be dynamically allocated is represented with type $\bRef$ in our encoding  to allow for  potential aliasing. 

\begin{example} \label{ex:trans-expr}
Suppose that contract $C$ has a state variable $\emph{\varn{x}}$ of Solidity type $\emph{\texttt{mapping(int => int[])}}$, which corresponds to the type $\emph{\texttt{int => (int => int)}}$ in our core Solidity language. 
The expression $\emph{\varn{x[0][1]}}$ will be translated into the Boogie expression
$M_{\bint}^{\bint}[e][1]$ where $e$ is $M_{\bint}^{\bRef}[~x^C[\this]~][0]$ using the rules from Figure~\ref{fig:enc-expr}.

\end{example}

\begin{figure*}[!t]
\centering
\[
\begin{array}{c}
\irulelabel
{\vdash e_1 \hookrightarrow \chi_1 \quad \vdash e_2 \hookrightarrow \chi_2}
{\vdash e_1 := e_2 \leadsto \chi_1 := \chi_2}
{\textrm{(Asgn)}}
~
\irulelabel
{\vdash e \hookrightarrow \chi}
{\vdash \solRequire(e) \leadsto \bassume \chi}
{\textrm{(Req)}}
~
\irulelabel
{\vdash e \hookrightarrow \chi}
{\vdash \solAssert(e) \leadsto \bassert \chi}
{\textrm{(Asrt)}} \\ \ \\

\irulelabel
{\vdash e \hookrightarrow \chi \quad \vdash s_1 \leadsto \omega_1 \quad \vdash s_2 \leadsto \omega_2}
{\vdash \solIfelse{e}{s_1}{s_2} \leadsto \bifelse{\chi}{\omega_1}{\omega_2}}
{\textrm{(Cond)}}
~
\irulelabel
{\vdash e \hookrightarrow \chi \quad \vdash s \leadsto \omega}
{\vdash \solWhile{e}{s} \leadsto \bwhile{\chi}{\omega}}
{\textrm{(Loop)}} \\ \ \\

\irulelabel
{\begin{array}{c}
    \vdash e_r \hookrightarrow \chi_r \quad \vdash e_i \hookrightarrow \chi_i \quad i = 0, \ldots, n \quad \textrm{fresh}~v \quad C_j <: \emph{Type}(\emph{this}) \quad j = 1, \ldots, m \\
    \begin{array}{rcl}
        \omega & \!\!\equiv\!\! & \bfont{if}~(\dtype[\this] = C_1)~ \set{\bcall{f^{C_1}}{\this, \chi_1, \ldots, \chi_n, \msgsender}{v};~ \chi_r := v }~ \bfont{else~if}~\ldots \\
               &                & \bfont{else~if}~(\dtype[\this] = C_m)~ \set{\bcall{f^{C_m}}{\this, \chi_1, \ldots, \chi_n, \msgsender}{v};~ \chi_r := v } \\
    \end{array}
\end{array}}
{\vdash e_r := f(e_1, \ldots, e_n) \leadsto \omega}
{\textrm{(IntCall)}} \\ \ \\

\irulelabel
{\begin{array}{c}
    \vdash e_r \hookrightarrow \chi_r \quad \vdash e_i \hookrightarrow \chi_i \quad i = 0, \ldots, n \quad \textrm{fresh}~v \quad C_j <: \emph{Type}(e_0) \quad j = 1, \ldots, m \\
    \begin{array}{rcl}
        \omega & \!\!\equiv\!\! & \bfont{if}~(\dtype[\chi_0] = C_1)~ \set{\bcall{f^{C_1}}{\chi_0, \chi_1, \ldots, \chi_n, \this}{v};~ \chi_r := v }~ \bfont{else~if}~\ldots \\
               &                & \bfont{else~if}~(\dtype[\chi_0] = C_m)~ \set{\bcall{f^{C_m}}{\chi_0, \chi_1, \ldots, \chi_n, \this}{v};~ \chi_r := v } \\
    \end{array}
\end{array}}
{\vdash e_r := e_0.f(e_1, \ldots, e_n) \leadsto \omega}
{\textrm{(ExtCall)}} \\ \ \\

\irulelabel
{\begin{array}{c}
    \vdash e_r \hookrightarrow \chi_r \quad \vdash e_i \hookrightarrow \chi_i \quad i = 1, \ldots, n \quad \textrm{fresh}~ v \\
    \omega \equiv \bcall{\malloc}{}{v};~ \bassume \dtype[v] = C;~ \bfont{call}~ f_{0}^{C}(v, \chi_1, \ldots, \chi_n, \this);~ \chi_r := v
\end{array}}
{\vdash e_r := \texttt{new}~C(e_1, \ldots, e_n) \leadsto \omega}
{\textrm{(NewCont)}} \\ \ \\

\irulelabel
{\vdash \varn{x}[\varn{x}.\length \texttt{++}] := e \leadsto \omega}
{\vdash \varn{x}.\push(e) \leadsto \omega}
{\textrm{(Push)}}
~
\irulelabel
{\begin{array}{c}
    \vdash e_r \hookrightarrow \chi_r \quad \vdash e \hookrightarrow \chi \quad \textrm{fresh}~v \quad \vdash v[i] \hookrightarrow \chi_i \\
    \omega \equiv \bcall{\malloc}{}{v};~ \bassign{\arrSel{\length}{v}}{\chi};~ \bassume{\forall i :: \chi_i = 0};~ \bassign{\chi_r}{v} \\
\end{array}}
{\vdash e_r := \texttt{new} \ t[](e) \leadsto \omega}
{\textrm{(NewArr)}} \\ \ \\

\irulelabel
{\begin{array}{c}
    \vdash s_1 \leadsto \omega_1 \\
    \vdash s_2 \leadsto \omega_2 \\
\end{array}}
{\vdash s_1; s_2 \leadsto \omega_1; \omega_2}
{\textrm{(Seq)}}
~
\irulelabel
{\begin{array}{c}
    \vdash e_r \hookrightarrow \chi_r \quad \textrm{fresh}~v \quad \vdash v[i_1]\ldots[i_n] \hookrightarrow \chi \\
    \omega \equiv \bcall{\malloc}{}{v};~ \bfont{call}~\textrm{MapInit}(v, n);~ \bassume \forall i_1, \ldots, i_n :: \chi = 0;~ \chi_r := v \\
\end{array}}
{\vdash e_r := \texttt{new}~ (t_1 \Rightarrow \ldots \Rightarrow t_n \Rightarrow t)() \leadsto \omega}
{\textrm{(NewMap)}} \\ \ \\

\end{array}
\]
\vspace{-20pt}
\caption{Inference rules for encoding Solidity statements to Boogie statements. $\emph{Type}(e)$ is a function that returns the static type of Solidity expression $e$. Symbol $f^C$ denotes the function $f$ in contract $C$, and $f_0^C$ denotes the constructor of contract $C$. The $<:$ relation represents the sub-typing relationship. $\dtype$ is a global Boogie map that maps receiver objects to their dynamic types. Types for universally quantified Boogie variables are omitted for brevity.}
\label{fig:enc-stmt}
\end{figure*}

\vspace{5pt}
\noindent \textbf{From Solidity to Boogie statements.}
Figure~\ref{fig:enc-stmt} presents the translation from Solidity to Boogie statements using  judgments of the form $\vdash s \leadsto \omega$ indicating that Solidity statement $s$ is translated to Boogie statement(s) $\omega$.
Since most rules in  Figure~\ref{fig:enc-stmt} are self-explanatory, we only explain our translation for assignments, function calls, and dynamic allocations.



\vspace{5pt}
\noindent \emph{Function calls.} Functions in Solidity have two implicit parameters, namely \varn{this} for the receiver object and \varn{msg.sender} for the Blockchain address of the caller. Thus, when translating Solidity  calls to their corresponding Boogie version, we explicitly pass these parameters in the Boogie version. However, recall that the value of the implicit \varn{msg.sender} parameter varies depending on whether the call is external or internal. For internal calls, \varn{msg.sender} remains unchanged, whereas for external calls,  \varn{msg.sender} becomes the current receiver object. For both types of calls, our translation introduces a conditional statement to deal with dynamic dispatch. Specifically, our Boogie encoding introduces  a map  $\dtype$ to store the dynamic type of receiver objects at allocation sites, and the translation of function calls invokes the correct version of the method based on the content of $\dtype$ for the receiver object.

\begin{figure}[!t]
\begin{lstlisting}[style=Boogie, numbers=none, basicstyle=\ttfamily \scriptsize]
var Alloc: [Ref]bool;
procedure New() returns (ret: Ref) {
  havoc ret; assume (!Alloc[ret]);
  Alloc[ret] := true;
}
procedure NewUnbounded() {
  var oldAlloc: [Ref]bool;
  oldAlloc := Alloc; havoc Alloc;
  assume $\forall i$::oldAlloc[i] ==> Alloc[i];
}
procedure MapInit(v: Ref, n: int) {
  var j: int; j := 1; Length[v] := 0;
  while (j < n) {
    assume $\forall i_1, \ldots, i_j$::Length[$\chi(v, i_1, \ldots, i_j)$]=0;
    assume $\forall i_1, \ldots, i_j$::$\neg$Alloc[$\chi(v, i_1, \ldots, i_j)$];
    call NewUnbounded();
    assume $\forall i_1, \ldots, i_j$::Alloc[$\chi(v, i_1, \ldots, i_j)$];
    assume $\forall i_1, \ldots, i_j, i'_j$::$(i_j = i'_j) \lor \chi(v, i_1, \ldots, i_j) \neq \chi(v, i_1, \ldots, i'_j)$;
    j := j + 1; }
}
\end{lstlisting}
\vspace{-10pt}
\caption{Auxiliary Boogie procedures. The term $\chi(v, i_1, \ldots, i_j)$ denotes the translation result of Solidity expression $v[i_1]\ldots[i_j]$. Types for universally quantified Boogie variables are omitted for brevity.}
\label{fig:aux-boogie-proc}
\vspace{-10pt}
\end{figure}

\vspace{5pt}
\noindent \emph{Dynamic allocation.}
Dynamic memory allocations in Solidity are translated into Boogie code with the aid of a helper  procedure $\malloc$ (shown in Figure~\ref{fig:aux-boogie-proc}) which always returns a fresh reference.
As shown in Figure~\ref{fig:aux-boogie-proc}, the $\malloc$ procedure is implemented using a global map $\alloc$ to indicate whether a reference is allocated or not. All three types of dynamic  memory allocation  (contract, array, and map) use this helper $\malloc$ procedure to generate Boogie code.

In the case of contract creation (labeled NewCont in Figure~\ref{fig:enc-stmt}), the Boogie code we generate updates the $\dtype$ map mentioned previously in addition to allocating new memory. Specifically, given the freshly allocated reference $v$ returned by $\malloc$, we initialize $\dtype[v]$ to be $C$ and also call $C$'s constructor as required by Solidity semantics.

Next, let us consider the allocation of array objects described in rule NewArr in Figure~\ref{fig:enc-stmt}. Recall that our Boogie encoding uses a map called \varn{Length} to keep track of the size of every array. Thus, in addition to allocating new memory, the translation of array allocation also updates the \varn{Length} array and initializes all elements in the array to be zero (or null).

Finally, the rule NewMap shows how to translate map allocations in Solidity to Boogie using an auxiliary Boogie procedure called \textrm{MapInit}  (shown in Figure~\ref{fig:aux-boogie-proc}) for map initialization.  Given an $n$-dimensional map, the \textrm{MapInit} procedure iteratively allocates lower dimensional maps and  ensures that  values stored in the map do not alias each other as well as any other previously allocated reference.

\begin{example} \label{ex:trans-stmt}
The Solidity code
\[
x := \texttt{new (int => int => int)()}
\]
is translated into the following Boogie code:
\begin{lstlisting}[style=Boogie, xleftmargin=5pt, numbersep=5pt, basicstyle=\ttfamily \scriptsize]
call v := New(); assume Length[v] = 0;
assume $\forall$i :: Length[$M_{\bint}^{\bRef}$[v][i]] = 0;
assume $\forall$i :: $\neg$Alloc[$M_{\bint}^{\bRef}$[v][i]];
call NewUnbounded();
assume $\forall$i :: Alloc[$M_{\bint}^{\bRef}$[v][i]];
assume $\forall$i,j :: i=j $\lor~ M_{\bint}^{\bRef}$[v][i] $\neq M_{\bint}^{\bRef}$[v][j];
assume $\forall$i,j :: $M_{\bint}^{\bint}$[$M_{\bint}^{\bRef}$[v][i]][j] = 0;
$x^C$[this] := v;
\end{lstlisting}
First of all, we allocate a fresh reference $v$ and initialize the length of $v$ and every inner map $v[i]$ to zero (lines 1 - 2). Second, we allocate fresh references for every inner map $v[i]$ (lines 3 - 5), and ensure the references of inner maps $v[i]$ and $v[j]$ do not alias if $i \neq j$ (line 6). Finally, we initialize every element $v[i][j]$ to zero and assign reference $v$ to the state variable $x$ (lines 7 - 8).
\end{example}

\subsection{Invariant Generation}

As mentioned earlier, translating Solidity code into Boogie allows \verisol{} to leverage the existing ecosystem around Boogie, including efficient verification condition generation~\cite{leino-ipl05}. However, in order to completely automate verification (even for loop and recursion-free contracts), we still need to infer a suitable contract invariant as discussed in Section~\ref{sec:verif-overview}. Specifically, recall that the contract invariant must satisfy the following two properties:
\begin{enumerate}
    \item $\models \hoare{\texttt{true}}{f_0}{\inv}$
    \item $\models \hoare{\inv}{f_i}{\inv} \text{~for all~} i \in [1, n]$
\end{enumerate}

\verisol{} uses monomial predicate abstraction (\cite{houdini,lahiri-cav09a,monomial})   to automatically infer contract invariants satisfying the above  properties.
Specifically, the contract invariant inference  algorithm conjectures the conjunction of all candidate predicates as an inductive invariant and progressively weakens it based on failure to prove a candidate predicate inductive.
This algorithm converges fairly fast even on large examples but relies on starting with a superset of necessary predicates. In the current  implementation of \verisol{}, we obtain candidate invariants by  instantiating the predicate template $e_1 \bowtie e_2$ where $\bowtie$ is either equality or disequality. Here,  expressions $e_1, e_2$ can be instantiated with variables corresponding to roles and states in the Workbench policy as well as constants. We have found these candidate predicates to be sufficiently general for automatically verifying semantic conformance of Workbench contracts; however, additional predicates may be required for other types of contracts.

\subsection{Bounded Model Checking}

If \verisol{} fails to verify contract correctness using monomial predicate abstraction, it employs an assertion-directed bounded verifier, namely \corral{}~\cite{lal-cav12}, to look for a transaction sequence leading to an assertion violation.
\corral{} analyzes the harness in Figure~\ref{fig:harness} by unrolling the loop $k$ times and uses a combination of {\it abstraction refinement} techniques (including lazy inlining of nested procedures) to look for  counterexamples in a scalable manner. 
Thus, when \verisol{} fails to verify the property, it either successfully finds a counterexample or verifies the lack of any counterexample with $k$ transactions.

\section{Evaluation}
\label{sec:eval}

\newcommand{\AssetTransfer}{\texttt{AssetTransfer}}
\newcommand{\BasicProvenance}{\texttt{BasicProvenance}}
\newcommand{\BazaarItem}{\texttt{BazaarItemListing}}
\newcommand{\DefectCompCounter}{\texttt{DefectCompCounter}}
\newcommand{\DigitalLocker}{\texttt{DigitalLocker}}
\newcommand{\FreqFlyerRewards}{\texttt{FreqFlyerRewards}}
\newcommand{\HelloBlockchain}{\texttt{HelloBlockchain}}
\newcommand{\PingPongGame}{\texttt{PingPongGame}}
\newcommand{\RefrigeratedTrans}{\texttt{RefrigTransport}}
\newcommand{\RoomThermostat}{\texttt{RoomThermostat}}
\newcommand{\SimpleMarket}{\texttt{SimpleMarketplace}}

We evaluate the effectiveness and efficiency of \verisol{} by performing two sets of experiments on smart contracts shipped with Workbench: (i) semantic conformance checking for  Workbench samples, and (ii) checking safety and security properties for the PoA governance contract. The goal of our evaluation is to answer the following research questions:
\begin{itemize}[leftmargin=25pt]
  \item[\textbf{RQ1}] How does \verisol{} perform when checking semantic conformance of Workbench application policies?
  \item[\textbf{RQ2}] How applicable is \verisol{} on smart contracts with complex data structures (such as PoA)?
\end{itemize}

\begin{table*}[t]
\centering
\vspace{5pt}
\caption{Experimental Results of Semantic Conformance against Workbench Application Policies.}
\begin{tabular}{|c|l|c|c|c|c|c|}
\hline
\multirow{2}{*}{\bf Name} &
\multirow{2}{*}{\bf Description} &
\multirow{1}{*}{\bf Orig} &
\multirow{1}{*}{\bf Inst} &
\multirow{2}{*}{\bf Init Status} &
\multirow{2}{*}{\bf Status after Fix} &
\multirow{2}{*}{\bf Time (s)} \\
 & & {\bf SLOC} & {\bf SLOC} & & & \\
\hline
\AssetTransfer & Selling high-value assets & 192 & 444 & Refuted & Fully Verified & 2.1 \\
\hline
\BasicProvenance & Keeping record of ownership & 43 & 95 & Fully Verified & Fully Verified & 1.5 \\
\hline
\BazaarItem & Multiple workflow scenario for selling items & 98 & 175 & Refuted & Fully Verified & 2.3 \\
\hline
\DefectCompCounter & Product counting using arrays for manufacturers & 31 & 68 & Fully Verified & Fully Verified & 1.3 \\
\hline
\DigitalLocker & Sharing digitally locked files & 129 & 260 & Refuted & Fully Verified & 1.7 \\
\hline
\FreqFlyerRewards & Calculating frequent flyer rewards using arrays & 47 & 90 & Fully Verified & Fully Verified & 1.3 \\
\hline
\HelloBlockchain & Request and response (Figure~\ref{fig:helloblockchain-fig}) & 32 & 78 & Fully Verified & Fully Verified & 1.3 \\
\hline
\PingPongGame & Multiple workflow for two-player games & 74 & 136 & Refuted & Fully Verified (manual)& 2.1 \\
\hline
\RefrigeratedTrans & Provenance scenario with IoT monitoring & 118 & 187 & Fully Verified & Fully Verified & 2.2 \\
\hline
\RoomThermostat & Thermostat installation and use & 42 & 99 & Fully Verified & Fully Verified & 1.3 \\
\hline
\SimpleMarket & Owner and buyer transactions & 62 & 118 & Fully Verified & Fully Verified & 1.4 \\
\hline
\hline
\textbf{Average} & - & 79 & 159 & - & - & 1.7 \\
\hline
\end{tabular}
\label{tab:samples-results}
\end{table*}

\vspace{5pt}
\noindent \textbf{Experimental Setup.}
Due to limited resources, we set our timeout as one hour for every benchmark.
All experiments are conducted on a machine with Intel Xeon(R) E5-1620 v3 CPU and 32GB of physical memory, running the Ubuntu 14.04 operating system.

\subsection{Semantic Conformance for Workbench Application Policies}
\label{sec:verisol-workbench}

\vspace{5pt}
\noindent \textbf{Benchmarks.}
We have collected all sample smart contracts that are shipped with Workbench and their corresponding application policies on the Github repository of Azure Blockchain~\cite{workbench-samples}.
These smart contracts and their policies depict various workflow scenarios that are representative in real-world enterprise use cases.
The smart contracts exercise various features of Solidity such as arrays, nested contract creation, external calls, enum types, and mutual-recursion. 
For each smart contract $\Contract$ and its application policy $\Policy$, we perform program instrumentation as explained in Section~\ref{sec:solidity-instrumentation} to obtain contract $\Contract'$. Note that no assertion failure of $\Contract'$ is equivalent to the semantic conformance between $\Contract$ and $\Policy$, so we include such instrumented smart contracts in our benchmark set.

\vspace{5pt}
\noindent \textbf{Main Results.}
Table~\ref{tab:samples-results} summarizes the results of our first experimental evaluation.
Here, the ``Name'' column gives the name of the contract, and the ``Description'' column  describes the contract's usage scenario. The next two columns give the number of lines of Solidity code before and after the instrumentation described in Section~\ref{sec:solidity-instrumentation}.
The last three columns present the main verification results: In particular, ``Init Status'' shows the result of applying \verisol{} on the original  smart contract, and ``Status after Fix'' presents the result of \verisol{} after we manually fix the bug (if any).
The fix may require changing either the policy or the contract, depending on the contract author's feedback.
Finally, ``Time'' shows the running time of \verisol{} in seconds when applied to the fixed contracts.

Our experimental results demonstrate that \verisol{} is useful for checking semantic conformance between Workbench contracts and the policies they are supposed to implement. In particular, \verisol{}  finds bugs in $4$ of $11$ well-tested contracts and precisely pinpoints the trace leading to the violation.
 Our results also demonstrate that \verisol{} can effectively automate semantic conformance proofs, as  it can  successfully verify all the contracts after fixing the original bug.
Moreover, for $10$ out of the $11$ contracts with the exception of \PingPongGame{}, the invariant inference techniques sufficed to make the proofs completely push-button.
Our candidate templates for contract invariant did not suffice for the \PingPongGame{} contract mainly due to the presence of  mutually recursive functions between two contracts.
This required us to manually provide a function summary for the mutually recursive procedures that states an invariant over the state variable $s_w$ of the sender contract represented by the $\texttt{msg.sender}$ (e.g. $s_w[\texttt{msg.sender}] = s_1 \vee s_w[\texttt{msg.sender}] = s_2$ where $s_i$ are states of the sender contract).
This illustrates that we can achieve the power of the underlying sound Boogie modular verification to perform non-trivial proofs with modest manual overhead. 
We are currently working on extending the templates for contract invariant inference to richer templates for inferring postconditions for recursive procedures.


\vspace{5pt}
\noindent \textbf{Bug Analysis.}
We report and analyze the five bugs that \verisol{} found in the Azure Blockchain Workbench sample contracts. These bugs can be categorized into two classes: (i) incorrect state transition, and (ii) incorrect initial state. We briefly discuss these two classes of bugs.

\begin{figure}[!t]
\centering
\begin{lstlisting}[style=Sol, numbers=none]
function Accept() public
{
  if (msg.sender != InstanceBuyer && $\label{line:at-line1}$
      msg.sender != InstanceOwner) {
     revert();
  }
  ...

  if (msg.sender == InstanceBuyer) { $\label{line:at-line2}$
     ...
  }
  else {
     // msg.sender has to be InstanceOwner
     // from the revert earlier
     if (State == StateType.NotionalAcceptance) { $\label{line:at-line3}$
        State = StateType.SellerAccepted;
     }
     else if (State == StateType.BuyerAccepted) { $\label{line:at-line4}$
        // NON-CONFORMANCE: JSON transitions
        // to StateType.SellerAccepted
        State = StateType.Accepted;         $\label{line:at-line5}$
     }
  }
  ...
}
\end{lstlisting}
\vspace{-0.1in}
\caption{Buggy function \texttt{Accept} of \AssetTransfer.}
\label{fig:assettransfer-smart-contract}
\end{figure}

\vspace{5pt}
\noindent \textit{Incorrect state transition.}
This class of bugs arises when the implementation of a function in the contract violates  the state transition stated by the policy.
\verisol{} has found such non-conformance in the \AssetTransfer{} and \PingPongGame{} contracts.
For instance, let us consider \AssetTransfer{}
\footnote{\url{https://github.com/Azure-Samples/blockchain/tree/master/blockchain-workbench/application-and-smart-contract-samples/asset-transfer}}
as a concrete example.
In this contract, actions are guarded by the membership of \texttt{msg.sender} within one of the roles or instance role variables (see Figure~\ref{fig:assettransfer-smart-contract}).
\verisol{} found the transition from state \texttt{BuyerAccepted} to state \texttt{Accepted} in the \texttt{Accept} function had no matching transitions in the  policy.
Specifically, the policy allows a transition from \texttt{BuyerAccepted}  to \texttt{SellerAccepted} when invoking the function \texttt{Accept} and \texttt{msg.sender} equals the instance role variable \texttt{InstanceOwner}.
However, the implementation of function \texttt{Accept} transitions to the state \texttt{Accepted} instead of \texttt{SellerAccepted}.
From the perspective of the bounded verifier, this is a fairly deep bug, as it requires at least 6 transactions to reach the state \texttt{BuyerAccepted} from the initial state.

\vspace{5pt}
\noindent \textit{Incorrect initial state.}
This class of bugs arises when the initial state of a smart contract is not established as instructed by the  corresponding  policy.
We have found such non-conformance in \DigitalLocker{} and \BazaarItem{}.
For instance, the  policy of \DigitalLocker{} requires the initial state of the smart contract to be \texttt{Requested}, but the implementation ends up incorrectly setting the initial state to \texttt{DocumentReview}.
In the \BazaarItem{} benchmark, the developer fails to set the initial state of the contract despite the policy requiring it to be set to \texttt{ItemAvailable}.



\subsection{Security Properties of PoA Governance Contract}
\label{sec:verisol-poa}

In this section, we discuss our experience applying \verisol{} to PoA governance contracts. We first give some background on PoA and then discuss experimental results.

\vspace{5pt}
\noindent \textbf{Background on PoA governance contracts.}
In addition to application samples, Workbench also ships a core smart contract that constitutes an integral part of the Workbench system stack.
PoA is  an alternative to the popular Proof-of-Work (PoW) consensus protocol for permissioned blockchains, which consist of a set of nodes running the protocol and validating transactions to be appended to a block that will be committed on the ledger~\cite{poa-cite}.
{\it Validators} belong to different organizations, where each organization is represented by an {\it administrator}.
The protocol for admin addition, removal, voting, and validator set management is implemented as the PoA governance contract.
It implements the Parity Ethereum's {\it ValidatorSet} contract interface and is distributed on the Azure Blockchain github~\cite{poa-github}.
The smart contract consists of five component contracts (\texttt{ValidatorSet}, \texttt{SimpleValidatorSet}, \texttt{AdminValidatorSet}, \texttt{Admin}, \texttt{AdminSet}) totaling around 600 lines of Solidity code.
The correctness of the PoA governance contract underpins the trust on Workbench as well the rest of Azure Blockchain offering.


The smart contract uses several features that make it a challenging benchmark for Solidity smart contract reasoning.
We outline some of them here:
\begin{itemize}[leftmargin=*]
\item The contracts use  multiple levels of inheritance since the top-level contract \texttt{AdminValidatorSet} derives from the contract \texttt{SimpleValidatorSet} which in turn derives from \texttt{ValidatorSet} interface.
\item It uses sophisticated access control using Solidity modifiers to restrict which users and contracts can alter states of different contracts.
\item The contracts maintain deeply nested mappings and arrays to store the set of validators for different admins.
\item The contracts use nested loops and procedures to iterate over the arrays and use arithmetic operations to reason about majority voting.
\end{itemize}

\vspace{5pt}
\noindent \textbf{Properties.}
We examined three key properties of the PoA contract:
\begin{itemize}
\item[{\bf P1:}] {\bf At least one admin}: The network starts out with a single admin at bootstrapping, but the set of admins should never become empty. If this property is violated, the entire network will enter into a frozen state where any subsequent transaction will revert.
\item[{\bf P2:}] {\bf Correctness of AdminSet}: The \texttt{AdminSet} is a contract that exposes a set interface to perform constant time operations such as lookup. Since Solidity does not permit enumerating all the keys in a mapping, the set is implemented as a combination of an array of members \texttt{addressList} and a Boolean mapping \texttt{inSet} to map the members to true.
The property checks the coupling between these two data structures --- (i) \texttt{addressList} has no repeated elements, (ii) $\texttt{inSet}[a]$ is true iff there is an index $j$ such that  $\texttt{addressList}[j] == a$.
\item[{\bf P3:}] {\bf Element removal}: Deleting an element from an array is a commonly used operation for PoA contracts.
PoA correctness relies on invoking this procedure only for an element that is a member of the array.
\end{itemize}

\vspace{5pt}
\noindent \textbf{Bugs found.} To check the three correctness properties (P1), (P2), (P3) described above, we first annotated the PoA governance contracts with appropriate assertions and then analyzed them using \verisol{}. In addition to uncovering  a previously known violation of the ``at least one admin'' property, \verisol{} identified a few other bugs that have been confirmed and fixed by the developers. In particular, \verisol{} found a bug that results in the violation of property (P3): When an admin issues a transaction to remove a validator $x$ from its list of validators, a call to event \texttt{InitiateChange} will be emitted after removing $x$ (using \texttt{deleteArrayElement}).
To persist the change, another call to \texttt{finalizeChange} is needed. However, the implementation actually allows two consecutive calls \texttt{InitiateChange} without a call to \texttt{finalizeChange}. As a result, this bug can result in the PoA contract to fail to remove validators that are initiated to be removed.


In addition to the manually-added assertions that check the three afore-mentioned properties, the PoA governance contracts contain additional assertions that were added by the original developers. Interestingly, \verisol{} also found violations of these original assertions. However, these assertion failures were due to developers mistakenly using \texttt{assert} instead of  \texttt{require}. Although both \texttt{require} and \texttt{assert} failures revert an execution, Solidity  recommends using \texttt{assert} only for violations of internal invariants that are not expected to fail at runtime. \verisol{} found five such instances of assertion misuse.





\vspace{5pt}
\noindent \textbf{Unbounded verification.}  Unlike the semantic conformance checking problem   for client contracts, verifying properties (P1), (P2), (P3) of the PoA contracts requires non-trivial quantified invariants and reasoning about deeply nested arrays. Thus, we attempted \emph{semi-automated} verification of the PoA contracts by manually coming up with contract/loop invariants and method pre- and post-conditions. In addition, inductive proof of some of the properties also requires introducing {\it ghost variables} that are not present in the original code. Fully automated verification of these properties in PoA governance contracts is an ambitious, yet exciting area for future work.

\section{Related Work}
\label{sec:related}

In this section, we discuss prior work on ensuring the safety and security of smart contracts.
Existing techniques for  smart contract security can be roughly categorized into various categories, including static  approaches for finding vulnerable patterns, formal verification techniques, and runtime checking.
In addition, there has been work on formalizing the semantics of EVM  in a formal language such as the $\mathcal{K}$ Framework~\cite{hildenbrandt-csf18}.
Finally, there are several works that discuss a survey and taxonomy of vulnerabilities in smart contracts~\cite{luu-ccs16,atzei-post17,nikolic-acsac18}.

\vspace{5pt} 
\noindent
\emph{Static analysis.} The set of static analysis tools are based on a choice of data-flow analysis or symbolic execution to find variants of known vulnerable patterns.
Such patterns include the use of reentrancy, transaction ordering dependencies, sending ether to unconstrained addresses that may lead to lost ether, use of block time-stamps, mishandled exceptions, calling \texttt{suicide} on an unconstrained address, etc. 
Tools based on symbolic execution include Oyente~\cite{luu-ccs16}, MAIAN~\cite{nikolic-acsac18},  Manticore~\cite{manticore-cite}, and Mythril++~\cite{mythril-cite}.
On the other hand, several data-flow based tools also exist such as Securify~\cite{tsankov-ccs18} and  Slither~\cite{slither-cite}.
Finally, the MadMax tool~\cite{grech-oopsla18} performs static analysis to find vulnerabilities related to out-of-gas exceptions. 
These tools neither check semantic conformance nor verify assertions.  Instead, they mostly find instances of known vulnerable patterns 
and do not provide any soundness or completeness guarantees.
On the other hand, \verisol{} does not reason about gas consumption since it analyzes Solidity code, and it also needs the vulnerabilities to be expressed as formal specifications. 

\vspace{5pt}
\noindent
\emph{Formal verification.}
F*~\cite{bhargavan-plas16} and Zeus~\cite{kalra-ndss2018} use formal verification for checking correctness of smart contracts.
These approaches translate Solidity to the formal verification languages of F* and LLVM respectively and then apply F*-based verifiers and constrained horn clause solvers to check the correctness of the translated program.
Although the F* based approach is fairly expressive, the tool only covers a small subset of Solidity without loops and requires substantial user guidance to discharge proofs of user-specified assertions.
The design of Zeus shares similarities with \verisol{} in that it translates Solidity to an intermediate language and uses SMT based solvers to discharge the verification problem. 
However, there are several differences in the capabilities of the two works. First, one of the key contributions of this paper is the semantic conformance checking problem for smart contracts, which Zeus does not address.
Second, unlike our formal treatment of the translation to Boogie, Zeus only provides an informal description of the translation to LLVM and does not define the memory  model in the presence of nested arrays and mappings.
Unfortunately, we were unable to obtain a copy of Zeus to try on our examples, making it difficult for us to perform an experimental comparison for discharging assertions in Solidity code.

\vspace{5pt}
\noindent
\emph{Other approaches.}
In addition to  static analyzers and formal verification tools, there are also other approaches that enforce safe reentrancy patterns at runtime by borrowing ideas from linearizability~\cite{grossman-popl18}.
Another work that is related to this paper is FSolidM~\cite{mavridou-post18}, which provides an approach to specify smart contracts using a finite state machine with actions written in Solidity.
Although there is a similarity in their state machine model with our Workbench policies,  they do not consider access control, and the actions do not have nested procedure calls or loops.
Finally, the FSolidM tool does not provide any static or dynamic verification support.

\section{Conclusion}
\label{sec:concl}

In this work, we described one of the first uses of automated formal verification for smart contracts in an industrial setting.
We provided formal semantics to the Workbench application configuration, and performed automatic program instrumentation to enforce such specifications. 
We described a new formal verification tool \verisol{} using the Boogie tool chain, and illustrated its application towards non-trivial smart contract verification and bug-finding.
For the immediate future, we are working on adding more features of the Solidity language that are used in common enterprise workflows and exploring more sophisticated  inference for inferring more complex contract invariants.



\bibliographystyle{plain}
\bibliography{refs}

\end{document}